\documentstyle[epsf]{mn}
\begin{document}
 
\title{The eclipsing X-ray pulsar X-7 in M33}
\author[G. Dubus {\it et al.}]
       {G. Dubus$^{1,2}$, P.~A. Charles$^{1}$, K.~S. Long$^{3}$, P.~J.
Hakala$^{4,5}$, E. Kuulkers$^{1}$\\
        $^1$ University of Oxford, Dept. of Astrophysics, Keble Road, Oxford OX13RH, UK\\
	$^2$ DARC, UPR 176 du CNRS, Observatoire de Paris Meudon, 5 place Janssen, 92195 Meudon, France\\ 
	$^3$ Space Telescope Science Institute, 3700 San Martin Dr., Baltimore MD 21218, US\\
	$^4$ Mullard Space Science Laboratory, University College London, Holmbury St. Mary, Dorking, Surrey RH56NT, UK\\
	$^5$ Observatory and Astrophysics Lab., University of Helsinki, PO Box 14, FIN-00014, Finland\\}
\date{May 1997}
 
\maketitle
\begin{abstract}
Using our extensive {\it ROSAT} X-ray observations of M33, we
confirm a 3.45 day eclipse period for the {\it Einstein} source X-7
\cite{larson} and discover evidence for a 0.31-s pulse period. The 
orbital period, pulse period and observed X-ray luminosity are remarkably
similar to SMC X-1. We therefore suggest M33 X-7 is a neutron star 
high mass X-ray binary with a 15-40 M$_\odot$ O/B companion and a 
binary separation of 25-33R$_\odot$ if the companion is almost filling its
Roche lobe.
\end{abstract}

\begin{keywords}
Galaxies:individual:M33  - Local Group - X-rays:stars - binaries:eclipsing 
\end{keywords}
\pagebreak

\section{Introduction}
M33 is an Sd spiral galaxy in the Local Group, which at $\sim 800$kpc 
\cite{vand},
is slightly more distant than M31.  At an inclination of $i\sim56$ \degr,
the disk of M33 is accessible to observation without a large amount of 
absorption. The first X-ray studies of
M33 made with the {\it Einstein} Observatory Imaging Proportional Counter
\cite{long} and High Resolution Imager \cite{markert,trinchieri}, resulted
in the detection of 13 sources brighter than $10^{38}$erg$s^{-1}$, one of
which, X-7, was flagged as variable. As a result of their luminosities
the X-ray sources in M33 are thought mostly to
to be supernova remnants or X-ray binaries, and X-7 clearly falls into
the latter category. Peres et al. \shortcite{peres}
and later, Schulman et al. \shortcite{schu2,schu3} discovered eclipse
episodes in X-7 and proposed a period of 1.78 days. This strongly suggests
that X-7 is a 
High Mass X-ray Binary (HMXB). A recent reanalysis of archival {\it
Einstein}, {\it ROSAT} and {\it ASCA} observations of X-7 by Larson and
Schulman \shortcite{larson} argues in favour of a 3.45 day period.

Here we discuss a considerably larger set of ROSAT observations of X-7 
which extend over a time base of 6 years. We use these observations to
confirm the 3.45 day period and report the discovery of a 0.31 s pulse
period, which allow us to establish the X-ray source as a pulsating
neutron star.  We have used this dataset previously to discuss the
temporal variability of the nuclear source X-8 \cite{dubus}
(hereafter paper I). A more general discussion of all of the sources 
will be the topic of a future publication.

\section{Observations}
Multiple observations using either the {\it ROSAT} High Resolution Imager
(HRI) or Position Sensitive Proportional Counter (PSPC) were carried out
between 1991 and 1997 (see Tr\"umper \shortcite{trum} for instrumental
details). Most of these were centered on the optical nucleus of M33 at
$\alpha~1^{\rm{h}}{33}^{\rm{m}}50\fs4$, $\delta~30^\circ39\arcmin36\arcsec$
(J2000). With a luminosity of $\approx 10^{38}$ erg/s \cite{long2}, X-7 is
easily detected in all cases. The different observations are presented in
table \ref{tb:obs}. The total exposure time is 233-ks of HRI (PSPC: 50-ks)
nucleus pointed observations and about 92-ks of off-centered HRI data. From
the long set of HRI observations pointed at the nucleus of M33, we get a
best position for X-7 of $\alpha(J2000)~1^{\rm{h}}{33}^{\rm{m}}34\fs2$ and
$\delta~30^\circ32\arcmin08\farcs7$, with an error of 5\arcsec ~to
10\arcsec.  This is based on deriving a correction for the ROSAT X-ray
position of the nucleus by using the accurately known optical position
\cite{devauc}.

The barycentre-corrected photon arrival times were extracted from a
1\arcmin~region centered on the best position for X-7. Background counts
were also extracted from a 2\arcmin~radius disk centered on
$\alpha~1^{\rm{h}}{33}^{\rm{m}}20\fs7$ and
$\delta~30^\circ31\arcmin51\farcs3$ (J2000), close to X-7 and at roughly the same
angle from the axis. This choice was dictated by the presence of another
source close enough to X-7 that a standard background annulus would have been
unreliable. Depending on the analysis method, we used the data both in
arrival time format and in spacecraft orbital averages. For the latter, we
group together good time intervals within 3-ks of each other ($\approx$
orbital time within which M33 is in the field of view) and correct for
background and vignetting as described in David et al. \shortcite{david}.
After corrections, the mean HRI count rates were about 11 counts ks$^{-1}$
(PSPC, 39 counts ks$^{-1}$) for X-7 and about 3.5 counts ks$^{-1}$ (PSPC,
1 count ks$^{-1}$) for the background. Deadtime corrections are negligible
at such low count rates.

\begin{table}
\caption{{\it ROSAT} observations of M33 X-7}
\begin{tabular}{lclc}
Obs. & Offset (\arcmin) & Dates & Duration (ks)\\
$rp23a$&  8 & 1991 29--30 Jul & 29.1\\
$rp23b$&  8 & 1992 10 Aug     & 5.0\\
$rp23c$&  8 & 1993 7--9 Jan   & 16.3\\
$rh20a$&  7 & 1992 8--12 Jan  & 19.1\\
$rh20b$&  7 & 1992 1--3 Aug   & 15.8\\
$rh86$& 12 & 1994 1--11 Aug  & 16.4\\
$rh87$&  8 & 1994 27 Jul -- 7 Aug & 30.5\\
$rh88$&  6 & 1994 27 Jul -- 7 Aug & 24.5\\
$rh89$&  8 & 1994 6--8 Aug   & 20.3\\
$rh60a$&  8 & 1994 10--11 Aug & 8.0\\
$rh46$&  8 & 1995 15--15 Jan & 24.6\\
$rh60b$&  8 & 1995 10--16 Jul & 40.9\\
$rh11n$&  8 & 1996 18 Jan -- 8 Feb & 46.4\\
$rh11a$&  8 & 1996 17--27 Jul & 44.6\\
$rh03$&  8 & 1997 10--14 Jan & 33.9\\
\end{tabular}

\medskip
The first three observations were made by the PSPC while the others were
carried out with the HRI.
Offset is the angle between X-7 and the pointing center (the nucleus of 
M33 in most cases). The abbreviations used correspond to the last digits 
of the actual sequence number.
\label{tb:obs}
\end{table}

\section{Analysis and Results}
The analysis methods are similar to the ones applied to the nuclear source
M33 X-8 described in paper I. We searched for power at frequencies between
$10^{-3}$ and $10$ Hz using a modified Fourier spectrum (see Appendix A).
On corrected mean orbital fluxes, we looked for periodicities using a 
Lomb--Scargle normalised periodogram.

\subsection{The 3.45 day eclipse period}
All of the observations of X-7 which included an eclipse feature were
flagged as variable with better than 99.9\% confidence. 
We then removed the data taken
during the eclipse times and found no evidence for variability, i.e.,
out of eclipse the source appears to be steady.  
\begin{figure}
\epsfxsize=8.5cm
\epsfbox{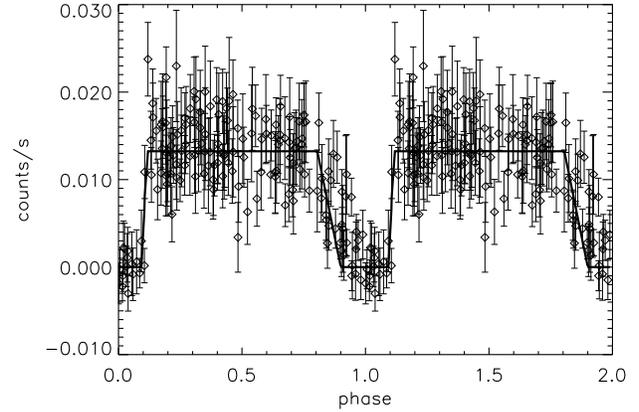}
\caption{Folded lightcurve of X-7 over the best period of 3.45d 
(two cycles are represented). Each point represents an orbital flux average.
The continuous line is the best fit model. Phase 0 corresponds to the 
mid-eclipse time derived in the text.}
\label{fig:period}
\end{figure}
We modeled the folded lightcurve from X-7 as a constant flux + linear
ingress and egress + an eclipse interval with zero flux. 
Minimization of the five
parameters (period P, mid-phase of the eclipse $\phi_{\rm mid}$, eclipse 
duration ${\Delta\phi}_{\rm eclipse}$, ingress duration 
${\Delta\phi}_{\rm ingress}$, egress duration ${\Delta\phi}_{\rm egress}$)
led to the following best fit ($\phi$=0 corresponds to JD 2448628.5) : 
{\vskip 0.05cm
\centerline{$P=298.38\pm0.02{\rm ~ks}$}
\centerline{$\phi_{\rm mid}=0.88\pm0.02~~~~{\Delta\phi}_{\rm eclipse}=0.20\pm0.03$}
\centerline{${\Delta\phi}_{\rm ingress}=0.10\pm0.05~~~{\Delta\phi}_{\rm egress}=0.01\pm0.01$}
\vskip 0.05cm}
The errors are estimated
from all the minima found with $\chi^2_\nu$$\approx$1
using different minimization routines,
initial values and number of free parameters. 
The period of 3.45 days is in agreement with that found by Larson et al.
\shortcite{larson}.
The values for 
$P$, $\phi_{\rm mid}$ and ${\Delta\phi}_{\rm eclipse}$ are quite robust
and accurately fitted. ${\Delta\phi}_{\rm egress}$ was always
close to 0 (symetric profiles led to a reduced $\chi^2_\nu$$\approx$2) 
while ${\Delta\phi}_{\rm ingress}$ was always $>$0. 
Using $P$, $\phi_{\rm mid}$ and the associated errors, we derive the following
ephemeris for the mid-eclipse time:
\centerline{JD$2448631.5\pm0.1+N(3.4535\pm0.0005)$}

The folded data with the model superimposed is shown in figure
\ref{fig:period}. The mean X-7 count rate in the constant part of the model
is found to be $\approx 13 {\rm ~counts} {\rm ~ks}^{-1}$.

\subsection{Evidence for a 0.31s pulsation} 
To search for periodicities at high frequencies, we split the PSPC 
and HRI data into 500s intervals of continuous data where X-7 is positively
detected (i.e. we exclude the eclipse intervals) and calculated the summed
Rayleigh power spectrum as described in the appendix. By doing so we lose 
about 
half of the available data to poor statistics. Smaller intervals would have 
resulted in less data being used because of the necessity to have at least
5 counts in each interval (see appendix A) while there are few 
continuous data intervals longer than 1000s. Choosing
an interval length of 1000s instead of 500s did not substantially change the
conclusions.

The resulting summed power 
spectrum is binned (reducing our frequency resolution)
to increase signal-to-noise. Assuming a $\chi^2$ distribution for the 
power (see appendix A),
logarithmic bins in frequency showed 
no deviations from Poisson statistics such as red noise in our range 
of 10$^{-3}$--10 Hz (photons are time-tagged to a resolution of $\sim$$10^{-3}$s). However, the data showed a significant (99.9\%
confidence) power enhancement at 3.17$\pm$0.08 Hz with linear 
binning, the error being the half-width of the peak in figure 2. 
The signal is washed out with logarithmic binning because of the 
corresponding poorer frequency resolution.
 Bin sizes of 70 to 120 independent Fourier
spacings gave similar results.

As discussed in the appendix, we relied on 1000 simulated Poisson fluxes
 to fully assess the significance of this peak.
The total number of counts $n$ in each simulated flux
is drawn from a Poisson distribution with the X-7 mean flux.
Using the properties of Poisson fluxes, the $n$ arrival times
are uniformly drawn between 0 and the exposure time, sorted
and then shifted to account for the observation
time window. The simulated data is processed as with X-7.
A Kolmogorov-Smirnov test comparing the power distribution of
the X-7 spectrum
with that calculated from the simulations showed no deviations, 
meaning the simulations adequately represented the flux from X-7. 
The maximum power at each frequency from the 1000 simulations gives
an estimate of the 99.9\% confidence level. As expected (see appendix), 
the peak in figure 2 is clearly above the maximum powers found in the 
simulations. This result is unchanged if Poisson fluxes are simulated 
with {\it exactly} the same number of counts as X-7 in each continuous
interval (to account for the eclipses).

Though the feature seems broader than what would be expected from
a simple sinusoidal pulse, it is not broad enough to
qualify as a quasi-periodic oscillation (see e.g. van der Klis, 1989) and
we therefore interpret it as an X-ray pulse. Knowing from the 
lightcurve the eclipse phase and orbital period, we 
tried correcting the photon times for their travel time 
across the system assuming a circular orbit. This did
not lead to conclusive results.
The Doppler effect and a non-zero pulse derivative are not enough 
to explain the observed peak width $\sim$0.1 Hz.
Assuming a $\sim$50 light-second binary separation by analogy with SMC X-1
(see discussion), 
the Doppler shift is only of the order of $10^{-3}$ Hz. A frequency drift of
the order of 0.1 Hz over $\sim$5 years implies $\dot{\nu}\sim
10^{-10}$ Hz s$^{-1}$, at least an order of magnitude 
higher than what is observed in galactic X-ray pulsars.

\begin{figure}
\epsfxsize=8.5cm
\epsfbox{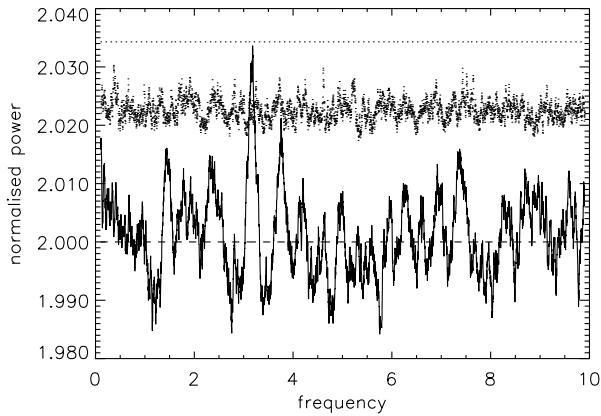}
\caption{Fourier power spectrum for X-7. The power spectra of 
500-s data intervals from the PSPC and HRI have been summed together.
The total power spectrum was then smoothed by averaging the nearest 100 
contiguous frequencies (there are 5000 independent frequencies in the 
non-averaged power spectrum). The dotted line is the theoretical 99.9\% 
confidence level assuming a $\chi^2$ statistic for the power 
distribution (the dashed line at y=2 being the expected average)
and the dots represent the maximum power found at each frequency 
in the 1000 simulated fluxes (i.e. the calculated 99.9\% confidence level 
see section 3.2).}
\label{fig:power}
\end{figure}

Our data spans $\sim$5 years, implying a folding frequency
resolution of $\sim 10^{-9}$Hz. 
As the count rate is not high enough to detect the
pulsation within each observation we cannot fold the 
data and give a pulse shape with confidence. This is not in
contradiction with the power spectrum analysis which does not include
phase information (i.e. the distribution of observations in time).
At this point it is not possible to check if the variability is coherent.
Although pulsed emission seems a reasonable assumption, the
power excess could arise from variability of a different nature (e.g.
broadband variability increasing the chance of spurious detection).

\section{Discussion}

\subsection{The orbital lightcurve}
Our well-sampled orbital lightcurve (figure \ref{fig:period}) shows clearly
that the X-7 eclipse ingress is gradual, but the egress is fast.
This is compatible with the standard picture of an extended region of the
disc which precedes the secondary in phase, analogous with features
well-known in low-mass X-ray binaries (see e.g. White {\it et al} 1995).  In
high inclination systems this extended disc region (created by the impact of
the mass transfer stream on the disc) will start to obscure the compact
object before its eclipse by the secondary.  We therefore believe that
Roche-lobe overflow plays a significant role in X-7, even if it is a HMXB. 

\subsection{System parameters}
As a 3.45-d eclipsing system with a 0.31-s pulse period, X-7 is very similar
to the high mass X-ray binary SMC X-1 which has an orbital period of 3.89-d,
a pulse period of 0.71-s and an X-ray luminosity of $5\cdot10^{38}$ erg
s$^{-1}$ \cite{paradijs}. The $10^{38}$ erg s$^{-1}$ luminosity of X-7 is
compatible with the Eddington luminosity of a neutron star accreting
at ${\sim}10^{-8}$M$_\odot$yr$^{-1}$. As is
the case with SMC X-1, the companion is likely to be an O or B star, 
as suggested by the location of X-7's
error box in the dense OB association HS13 \cite{humphreys}. 
Roche lobe overflow is also significant in SMC X-1 \cite{paradijs}, and so
we shall assume in our calculations that the mass-losing star in X-7 either
fills or is close to filling its Roche lobe.

Under this assumption, we take $R_{\rm Roche}=R_2$ to be the radius of the
star, and adopt a binary inclination of 90$^\circ$.  We can now use our
observed eclipse duration of 0.20 in phase to give $R_2/a = 0.59$, where $a$
is the binary separation. Now the ratio of the Roche lobe radius to $a$ is
(for circular orbits) a function only of the mass ratio $q$ (=$M_X/M_2$) of the
binary \cite{eggleton}, from which we derive $q = 0.085$.  Combining this
with Kepler's third law and an assumed compact object mass of 1.4M$_\odot$
(a canonical neutron star mass for our observed X-ray pulsar) gives $a =
25$R$_\odot$, $M_2 = 16.4$M$_\odot$ and $R_2 = 14.8$R$_\odot$. This
corresponds to an early-type star, and is significantly larger than a ZAMS
star (which would have a radius of only 5.8R$_\odot$ for this mass). These
parameters are consistent with a B0III star which has M$_V$ = -5.1.

However, the inclination could be as low as 70$^\circ$, for which our data
would then give $q = 0.036$, $a = 32.9$R$_\odot$, $M_2 = 39$M$_\odot$ and
$R_2 = 21$R$_\odot$.  Such a star is closer to a somewhat brighter (M$_V$ =
-6.5) O7I spectral type, and gives an indication of the uncertainties
involved in this calculation.  Interestingly, the brightest star in HS13 has
V = 18.8 and, with a distance modulus of 24.65 \cite{humphreys}, this
corresponds to an absolute magnitude mid-way between the B0III and O7I
spectral types.

\section{Conclusion}
We confirm M33 X-7 has an eclipse period of 3.45 days as Larson \& Schulman
\shortcite{larson} had found using data from different satellites. Using a
Fourier power spectrum method adapted to low count rates, we also find
evidence for a 0.31-s period, which identifies the compact object as a
neutron star. From our best fit to the orbital lightcurve, we find the
companion is an early type  O or B star with $M\geq$15M$_\odot$ and which is
likely to be
losing mass predominantly by Roche lobe overflow.
The optical location of X-7 within a dense O-B association is consistent
 with the expected massive companion. When identified, the
optical counterpart is likely to show ellipsoidal and/or heating variations
\cite{paradijs}.

\section*{Acknowledgments}
We would like to thank D. Pelat, J.-L. Masnou, S. Bonazzola, M. van der Klis and the referee for helpful comments on this paper. We also aknowledge help from E. Schulman in obtaining all the archival observations of M33.
We acknowledge support by the British-French joint research
programme {\it Alliance}, by NASA grant NAG 5-1539 to the STScI
and by the Academy of Finland.
This research has made use of data obtained through the High Energy
Astrophysics Science Archive Research Center Online Service, provided by the
NASA-Goddard Space Flight Center.

\protect

\appendix
\section{Time analysis at high frequencies}

\subsection{Motivation}
Searching for periodicities in our observations with the Lomb-Scargle periodogram proved impractical at high frequencies $>$ 0.001 Hz (the timescale of an orbit). First, the periodogram only extends to the Nyquist frequency of the data which is inversely proportional to the bin size chosen for the time series. The flux from X-7 being rather weak, small bin sizes imply long time series of zeros with a few 1's at the binned photon arrival times. This can be computationally problematic to handle. Second, the folding frequency
resolution of $\sim 10^{-9}$Hz for our dataset leads to an enormous number
of frequencies to search above 0.001 Hz.

In a Fourier analysis, it is thus customary (e.g. van der Klis, 1989) to split long observations into S smaller continuous intervals of {\it equal length}, 
bin the photon arrival times on some multiple of the satellite clock and then sum the resulting renormalised \cite{leahy} Fast Fourier Transforms (FFT). 
With the Leahy renormalisation, the power density probability follows a $\chi^{2}_{2{\rm S}}$ law when the number of counts in each interval is sufficiently large. Splitting observations increases the signal-to-noise ratio as compared to a single interval power spectrum by a factor $\sqrt {\rm S}$, though at the price of frequency resolution.

\subsection{Rayleigh power}
Binning makes no sense with a small number of events so, instead of the Leahy power spectrum, we calculated a Rayleigh power spectrum for each interval :
\[
{n \over 2} F_n(\nu)={\left({\sum^n_{k=1} \cos(2\pi\nu t_k)}\right)}^2+{\left({\sum^n_{k=1} \sin(2\pi\nu t_k)}\right)}^2
\]
if $n$ photons were recorded at times $t_k$ in the interval $[0,T]$ and $\nu$ is one of the independent frequencies $\{1/T,2/T,...\}$. The Rayleigh power is essentially the Fourier power of the photon arrival times modeled as diracs at $t_k$ (infinitely small bin size). The only limits to the highest sampled frequency are the time-tagging resolution of the instrument or the computation time requirements. As above, the power density probability follows a $\chi^{2}_{2}$ law when the number of counts in the interval is sufficiently large ($n\ga$ 100; see Brazier, 1994, and references therein).

However, the probability law is already quite close to a $\chi^{2}_{2}$ when $n\ga 5$ though the tail of the distributions statistically differ. In practice, simulations show that the probability to have a large value for $F_n(\nu)$ when $5 \la n \la 100$ is smaller than the corresponding probability for a $\chi^2_2$ distribution i.e. the significance of a high peak value in the power spectrum will be under-estimated if it is calculated with a $\chi^2_2$ distribution. If one assumes the $\chi^2_2$ is still an adequate representation of the power probability law even with low $n$, then the techniques described by van der Klis (1989) for the Leahy renormalised power spectra can be similarly applied (summed intervals, frequency binning). If a signal is detected (the threshold will in effect be more stringent than if $n$ is large), the actual significance has to be estimated through simulations (see section 3.2). 
\end{document}